# Latest Progresses in Developing Wearable Monitoring and Therapy Systems for Managing Chronic Diseases

## Han Gui[1] and Jing Liu[1, 2*]

1. Department of Biomedical Engineering, School of Medicine,
Tsinghua University, Beijing 100084, P. R. China

2. Technical Institute of Physics and Chemistry, Chinese Academy of Sciences,
Beijing 100190, P. R. China

* E-mail address: jliu@mail.ipc.ac.cn; Tel. +86-10-82543765

**Abstract**

Chronic disease is a long-lasting condition that can be controlled but not cured. Owing to the portability and continuous action, wearable devices are regarded as promising ways for management and rehabilitation of patients with chronic diseases. Although the monitoring functions of mobile health care are in full swing, the research direction along wearable therapy devices has not been paid with enough attention. The aim of this review paper is to summarize recent developments in the field of wearable devices, with particular focus on specific interpretations of therapy for chronic diseases including diabetes, heart disease, Parkinson's disease, chronic kidney disease and movement disability after neurologic injuries. A brief summary on the key enabling technologies allowing for the implementation of wearable monitoring and therapy devices is given, such as miniaturization of electronic circuits, data analysis techniques, telecommunication strategy and physical therapy etc. For future developments, research directions worth of pursuing are outlined.

## 1. Introduction

With the progresses in science and technology, residents in industrialized cities are now living longer, but with multiple complicated health conditions [1]. Clinical operations are also increasingly accurate, however, prevalence of chronic disease and aging population begin to trouble the healthcare system accompanying rising costs [2,3]. Chronic diseases like diabetes, high blood pressure, Parkinson's disease and so forth have limited treatment methods due to complicated and confused causes while narrow part of pathologic phenomenon and biological mechanisms have been approved in current biological sciences [4-6]. Surgery might be a solution to certain chronic diseases, for example, implanting electrodes in the brain of Parkinson's disease patient to control tremble and dyskinesia [7]. However, the high cost of surgical operations and large risk of postoperative infection overwhelm the elderly patients who are the main group of chronic diseases. The most frequent therapies are regular drug and healthy behaviors of diet and exercise which contribute to relieving the incurable burden and improving health outcome and quality of life [8,9]. Clearly, the optimal therapy for chronic diseases will be complex and require upgrading the health care system. However, part of the solutions may lie in how we take advantage of current advances in information technology to develop efficient system that is required to cope with the increased life expectancy.

The conception of mobile health appears on the scene to relieve the pressure of clinic management and to intervene the reinforcement of preventive medicine [10]. There have been many breakthroughs in monitoring system using mobile devices like mobile phone and personal digital assistant, which could implement electrocardiogram monitoring, diabetes control and dermatosis care and more [11-13]. On this sound foundation, an emerging research trend of mobile health places more importance on developing therapy technique which is the ultimate goal of wearable devices. As the developments of wireless communication and data analysis technologies, continuous transmission of patients' health status information to remote center can receive intervention from physicians and acquire complete self-management to chronic diseases [14,15]. Also, the approach of physical therapy including electrical stimulation, phototherapy, ultrasound and so on can be used to intervene the body, focus on musculoskeletal, neuromuscular, cardiovascular, pulmonary and integumentary disorders [16]. Besides, the miniaturization of equipment and the development of circulation system based on existing heavy therapeutic devices open a door to wearable therapy devices. Fig.1 showed the wearable monitoring and therapy devices throughout the body via the methods mentioned above, which feature long-term functional treatment to one or more kinds of chronic diseases, exploding into a prominence in the field of health care in the last few decades.

In this context, the present article aims to present a comprehensive review of the latest wearable devices for monitoring and therapy. In the following section, key technologies will be discussed which enabled the development and deployment of wearable devices for monitoring and therapy. The next two sections will demonstrate the examples of application of these technologies, while advantages and disadvantages of the specific systems will also be mentioned. Finally, current development and challenges will be outlined, as well as a brief outlook on future breakthroughs.

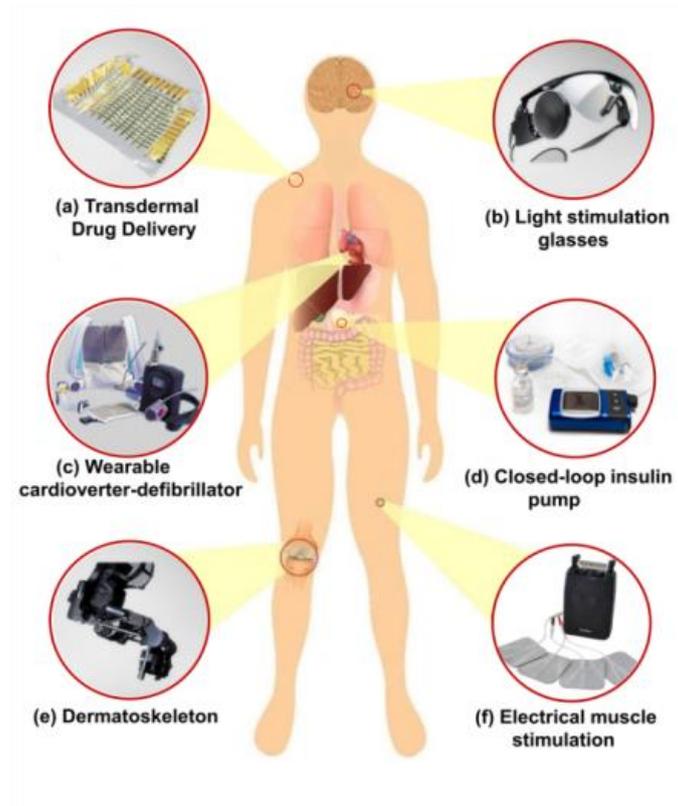

Fig. 1 Applications on wearable monitoring and therapy devices for chronic diseases

## 2. Key enabling technologies

Wearable devices for patients' monitoring and therapy means applying the portable and wearable systems to the medical field, including physiological and biochemical monitoring out of hospital, movement assistance, precise drug delivery, physical therapy and more. Recent advances in sensor technology, microelectronics, microelectromechanical system (MEMS), data analysis technology, telecommunication and physical therapy have enabled the development of wearable systems for patients' monitoring and therapy.

The miniaturization of electronic circuits based on the use of microelectronics has played an important role on the wearable devices. Looking back to history, it is not surprised to find how to downsize the medical equipment via a gradual process. Using electrocardiogram (ECG) as an example, Fig. 2 illustrates the evolution of devices from water buckets [183], bench-top [184] and portable sizes [185] to the recent wearable devices like belt, wrist strap and clothing [17,186,187], which may evolve into flexible and stretchable film with organic electronics in the next stage [18]. One conception closely relevant to wearable devices is the continuous monitoring and therapy. In the past, the size of sensors and front-end electronics made the hardware gathering physiological and biochemical data too obtrusive to be suitable for long-term monitoring application. Now developments in the field of microelectronics have allowed researchers to develop miniaturized circuits possessing sensing capability, microcontroller functions and radio transmission. MEMS has also enabled the development of miniaturization, which can realize batch fabrication reducing the cost of electronic components significantly. In addition, MEMS is the foundation of microneedle for transcutaneous detection and drug delivery. The painless and precise characteristics promise microneedle large prospects in the application of wearable monitoring and therapy devices.

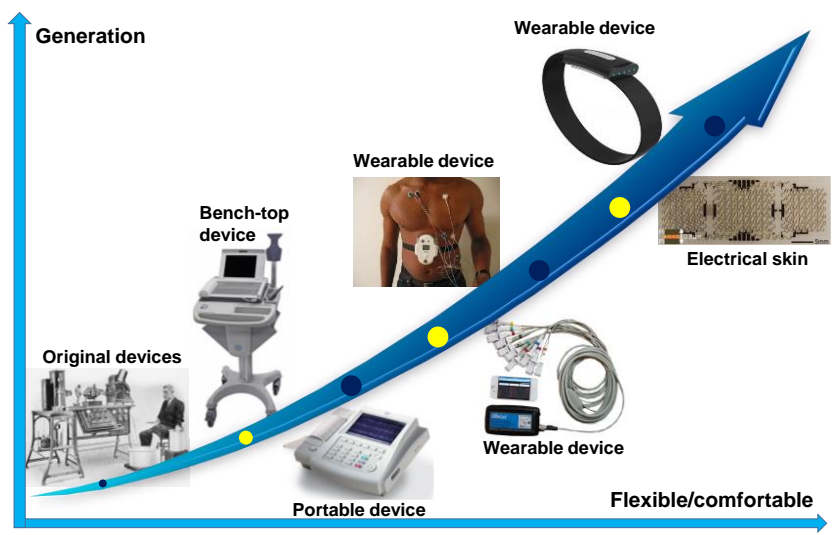

Fig. 2 Evolution of ECG devices with the miniaturization of electronic circuits [17,18,201-205]

Wearable systems for monitoring and therapy often integrate multiple sensors into a network which is coordinated with other actuator and processor units by wireless communication. Conducting this kind of short-range information transmission inside the wearable device, Bluetooth is a preferable choice. As for the long-range data transmission to remote severs, an information gateway, which is a personal device such as smart phone or personal computer with powerful processors and large memory capacity, is needed to gather the physiological signals from the sensor network, preprocess the received data and retransmit them to a central server. Advances in wireless communication have enabled the remote transmission of data in a short time. There is a wide variety of available wireless technologies that can serve this goal. Such technologies including WLAN, WiMAX as well as mobile phone communication standards like 4G can offer wide coverage and ubiquitous network access.

Data management is the bridge connecting complex human bodies' parameters and specific healthcare services. Automatic integration of collected data or manual input into research database can provide the medical community or the caregiver with an opportunity to prediction of personal health trends, early detection of disorders, enabled insight into illness evolution, the effects of medicine therapy and the rehabilitation process, which help patients suffering chronic disease more likely to feel in control of their health as well. System may not rely on the active participation and intervention of the user when the closed loop between sensor network and actuators like pump and electrode carry out the therapy function directly. One of the major data obstacles to the adoption of wearable device is the massive amount of data during the long-term and multiple monitoring of patients' status out of hospital. Data analysis techniques such as signal processing, pattern recognition, data mining and other artificial intelligence-based methodologies have enabled the continuous monitoring that would have been otherwise impossible. Besides, effective algorithm to control the dynamic physiological parameters and deal with emergencies are also important issues. For example, smart algorithms for the closed loop insulin pump are urgently needed to accomplish the proper medicine titration in various occasions including exercise, diet and overnight avoiding time lags in subcutaneous glucose sensing and insulin action [19].

Finally, a discussion of physical therapy for wearable device in this review paper mainly refers to two aspects: functional trainings and physical agent therapies including electrotherapy, phototherapy, ultrasound therapy, magnetotherapy, compression therapy and so on. Functional trainings combined with wearable devices could provide physically disabled individuals due to neurological disorders with high intensity and repetitive training and augmented feedback to improve performance [20]. It is now widely accepted that the brain has the potential for neural repair following stroke and other injuries through neural plasticity [21]. Positive rehabilitation through neuroplasticity depends on the functional tasks and the number of repetitions both of which the wearable therapy systems can meet in our review. With regard to physical agent therapies, the most comprehensive applied approach is electrotherapy. Direct current therapy is also called transdermal iontophoresis, which uses an electric field to move both charged and uncharged drugs across the skin [22]. Also, reverse iontophoresis extracting glucose from the skin is applied in the continuous glucose monitoring and provide a foothold for artificial pancreas. Another electrotherapy discussed in this review is functional electrical stimulation (FES) that stimulates the patient's muscles by previously configured electrical impulses to generate joint movement as a means of gait compensation. In general, although there is not thorough and systematic cognition on the therapeutic mechanism of wearable devices, it remains a potentially useful remedy to seek for rehabilitation scheme from the basic physical methods and biochemical effects following up the current technologies.

## 3. Wearable sensors for physiological parameters monitoring

Physiological measures for health monitoring include body temperature, respiration rate, heart rate, electrocardiogram (ECG), electroencephalogram (EEG), blood pressure (BP), blood glucose, blood oxygen saturation ($SpO_2$), electromyography (EMG) and muscle activity. Until recently, continuous monitoring of physiological parameters can only be realized in the hospital setting. However, the possibility of accurate, continuous and long-term monitoring of physiological parameters out of hospital is a reality with developments in the field of wearable technology. Here, we will illustrate some wearable sensors for different types of physiological parameters monitoring and list out their respective characteristics and issues.

### 3.1 Bioelectrical signal monitoring

The activities of human organs, such as heart, muscle, brain and so on, result in bioelectric signals which are defined as electric potentials between points in living cells and can be measured with several techniques.

ECG is one of the most widely used bioelectrical signal and many wearable ECG systems have been available in the market to date. A wearable physiological monitoring system for space and terrestrial applications named "Life Guard" to monitor the health status of astronauts in space was developed by Stanford University and NASA [17], where ECG signals and heart rate were acquired with traditional button electrodes providing ECG Lead II and V5. Thereafter the spread of mobile phone and the GSM (global system for mobile communication) mobile telephony network pushed the wearable ECG device with real-time telemonitoring to a new stage [23], which could help out-patients record their conditions for clinical analysis, emergency detection and effective rescue [11,24,25]. To get broader wearable application and better user experience, the electrode of ECG may deserve a discussion. Conventional wet ECG sensor always accounted for the complete market for medical use [26] but the drawbacks were also obvious that the pastes or gels to keep the electrical contact could cause skin irritation and discomfort. Especially when sweating in rigorous physical exercise, the electrodes might become loose and break electrical contact. Therefore, new dry-contact and noncontact electrodes through biopotential sensor technologies, using the skin and the electrode as the two layer of a capacitor, were researched to address these problems. Gruetzmann et al. [27] demonstrated a foam electrode which exhibited excellent stability with increased resistance to motion artifact versus the wet electrode. Park et al. [28] developed a wearable capacitive ECG monitoring system that did not require direct contact to the skin but possessed comparable performance to gold standard conventional electrodes. More typical nonadherent capacitive ECG sensing devices are summarized in Table 1. They could not only be worn by the subjects in the form of clothing, belts, watches and shoes, but also be embedded into ambient environment such as mattress or back of chair, bed, steering wheel of car and more. The major challenges of wearable ECG devices with non-adherent are the capacitive mismatch due to motion artifacts and the high contact impedance due to the indirect contact [29].

EMG is a good measure of the strength of contraction of the muscle, which can be used as a diagnostics tool for identifying neuromuscular diseases, or as a control signal for prosthetic devices such as prosthetic hands, arms and lower limbs. There are two kinds of EMG: surface EMG and intramuscular EMG. Surface EMG is more commonly used for wearable monitoring. Though there is some variation in surface EMG compared with original EMG, it can response the pattern of muscle contraction to a certain extent and provided a lot of information for action recognition [30]. As EMG signals contain some undesired sources like ECG artifacts, filtering techniques is suggested to eliminate all the noises. However, this process may reduce the noises but the quality of EMG signals is not guaranteed. For real-time monitoring, the classification of EMG signals is an important issue [31]. Accurate hand gesture classification including the position and pressure of finger presses, as well as the tapping and lifting gestures across has been realized by using forearm electromyography [32, 33]. Anna et al. [34] presented the design of a wearable device for reading positive expressions from facial EMG signals. Due to the intrinsic issues of electrode placement, a multi-attribute decision-making was adapted to find adequate electrode positions on the side of face to capture these signals. Applications of this wearable sensor can be extended to the medical support by long-term facial expression recognition during therapeutic interventions. Besides, the combination of SEMG and ACC signals was demonstrated to have the superiority both fine subtle actions and large-scale movements [35]. Further, the identification of highly discriminative features has been realized by extracting information from the EMG signals for the automatic assistance of people instability, which could be applied in continuous daily activity monitoring and fall detection [36, 37].

EEG is a non-invasive method for measuring brain activity. Compared with EMG control for body limbs, EEG-based brain-computer interfaces (BCI) translate brain activity into control signals without physical movement, which can be used as a means of communication for individuals with severe disabilities [49]. Though the clinical potential of enabling communication through an EEG BCI remains unclear, many key issues including the classification algorithms [50], the acquisition of BCI control [51], psychological principles of BCI use [52], and wireless and wearable EEG system design [53] have been discussed. Besides BCI communication, EEG monitoring can be applied to automated diagnosis of epilepsy [54-56], Alzheimer's disease [57], and even sleep disordered breathing [58]. Selecting features that best describe the behavior of EEG signals are important for automated seizure detection performance. Many time-domain [59, 60], frequency-domain [61-66], time-frequency analysis [67], energy distribution in the time-frequency plane [68, 69], wavelet features [62, 70] and chaotic features such as entropies [60, 71] are used for seizure detection.

Table 1 Non-adherent capacitive ECG sensing devices

| Devices | Location of ECG electrodes | Measured signals | References | Year |
|---|---|---|---|---|
| ECG Monitoring System | Attached on T-shirts | ECG | [28] | 2006 |
| ECG of neonates and infants | Attached on underwear | ECG | [38] | 2006 |
| Wearable smart shirt | Attached on shirt with wireless network | ECG, heart rate, Physical activity | [39] | 2009 |
| Wearable ECG sensor | Integrated into a cutton T-shirt | ECG | [40] | 2012 |
| Ambulatory ECG monitoring | Integrated on underwear | ECG | [40] | 2007 |
| Non-contact ECG/ EEG electrodes | Integrated into fabric and clothing | ECG, EEG | [42] | 2010 |
| ECG measurement in clinical practice | Integrated into a pillow of a pre-ambulance bed | ECG | [43] | 2009 |
| Sleeping bed | Between the mattress and bed sheet | ECG, heart rate, blood pressure | [44] | 2009 |
| Aachen smart-chair | Integrated into the backrest and pad of a chair | ECG | [45] | 2007 |
| Non-contact chair based system | Integrated into the backrest of a chair | ECG, BCG, heart rate, blood pressure | [46] | 2012 |
| Non-contacting ECG measurement | Attached on toilet seat | ECG | [47] | 2004 |
| Wireless steering wheel | Integrated into steering wheel | ECG, heart rate | [48] | 2012 |

## 3.2 Acoustic signal monitoring

Physiological acoustic signals from the human body include heart sounds, respiratory sounds, gastrointestinal sounds, limb joint sounds and more. Continuous monitoring and quantitative analysis of these physiological sounds are expected to play important role in the emerging wearable and mobile healthcare field. Here, two kinds of acoustic sensors were discussed.

The first introduced sensor was microphone, which can be subdivided into electret, MEMS

and piezoelectric film microphones. Corbishley et al. [72] demonstrated a miniaturized and wearable system for respiratory rate monitoring. The microphone was placed on the neck to record acoustic signals associated with breathing, which were band-pass filtered to obtain the signal modulation envelope. By developing techniques to filter the noise sources and interference artifacts, the average success in the measurement of breathing rate had achieved 91.3%. On the basis of breathing rate monitoring technology, heart rate could also be extracted by placing a small sensor at the neck [73]. An algorithm for the segmentation of heart sounds (S1 and S2) and extraction of heart rate was presented to avoid the weakness of conventional method of acoustic signal acquisition that needed to place the sound sensor at the chest where this sound is most audible. The authors managed to achieve accuracy greater than 90% with respect to heart rate value provided by two commercial devices. Moreover, the sensing methods of joint acoustical emissions from the knee were also proposed in many researches [74-76]. Friction between the structures and articulating components of the knee joint gives rise to various kinds of vibrations, which can be continuously monitored by miniature microphones places on the knee for wearable rehabilitation assessment. Changes in joint sounds during recovery from musculoskeletal injury could be quantified by analyzing the consistency of the knee acoustical emission during different activities [75].

Another kind of acoustic senor discussed in this article was accelerometer, which could detect the vibration of the skin due to sounds generated inside human body [77]. Typical applications of acoustic accelerometer sensors covered vocal behavior monitoring [78], heart sound monitoring [79], respiratory sound monitoring [80], fetal monitoring [81], and even gastrointestinal sound monitoring [82]. Compared with microphones, accelerometers are more sensitive to motion noise. Some algorithms have been developed to remove motion artifacts. For example, Pandia et al. [83] provided a method to effectively extract heart sound signals from accelerometer data overwhelmed by motion artifacts. Besides, friction noise generated between the sensor and the skin or clothes is also a big challenge. There a lot of methods to reduce or prevent friction noise including fixing the sensors to cloth, using low friction wires and slippery film coatings on the sensor surface and more.

## 3.3 Optical signal monitoring

Due there is not light source in human body, here we focused on the photoplethysmographic (PPG) sensing that involves a light-emitting diode to emit light into tissue and a photodiode to measure the amount of light reflected from or transmitted through the tissue. Because blood flow to the skin can be modulated by multiple physiological systems, the PPG can be used to monitor $SpO_2$ [84], heart rate [86], respiration rate [86] and BP [87].

In recent researches, the PPG sensors have been explored to design compact and lightweight for wearable measurement, which can be placed in different sites including the finger, wrist, brachia, earlobe, external ear cartilage, and the superior auricular region. Commercial clinical PPG sensors commonly use the finger, earlobe and forehead [88]. Earlobe sensors have become popular as pulse rate monitors because the transmitted PPG signal amplitude from the earlobe provides the largest perfusion value. Patterson et al. [89] introduced an ear-worn and flexible PPG sensor for heart rate monitoring, which is suited for long-term and continuous monitoring due to its location and unobtrusive design. However, over the course of a variety of daily activities like walking or jogging, the PPG sensor signal may become contaminated with motion artifacts. Relatively, sensors based on finger measurement sites are more easily to access and acquire good signal. For example, to minimize motion artifacts, a double ring design was developed to reduce the influence of external forces, acceleration and ambient light, and to hold the sensor securely to the skin, so that the blood circulation in the finger remained unobstructed [90-91]. Xu et al. [92] developed a near-infrared phototransistor with high sensitivity based on organic bulk heterojunction, which showed great potential for the design of low power PPG sensor. This phototransistor can be used as the detection unit in a PPG sensor for continuous finger pulse wave monitoring. Besides, novel PPG sensors with the wristwatch-type design have also been developed. Unlike previous methods that obtained signals from the capillaries in the skin, this method measures the signal from the radial artery and the ulnar artery of the wrist. Nemati et al. [93] proposed a promising wrist watch with PPG sensors for atrial fibrillation monitoring, which is the first study of enough high accuracy for general population without an ambulatory Holter electrocardiographic monitor. Lastly, forehead sensors have shown greater sensitivity to pulsatile signal changes under low perfusion conditions, compared with other peripheral body locations [94]. The forehead structure including the thin-skin layer and the prominent bone helps to direct light back to the photo detector, showing the result of decreased motion artifacts during different physical activities.

## 3.4 Biochemistry index monitoring

Biochemical index monitoring has recently gained a great deal of interesting among researches in the field of wearable technology. Combined this article's subject of chronic disease management, we focus the discussion on continuous glucose monitoring (CGM), which measures interstitial glucose as a marker of changes in blood glucose concentration, has more potential benefits than capillary blood glucose with point sample in the management of patients with type 1 diabetes [95].

The technology of CGM was introduced 15 years ago when some devices were approved by the U.S. Food and Drug Administration (FDA) and CE in Europe [96] for use, based on invasive subcutaneous sensor, minimally invasive microneedle and noninvasive methods mainly including transdermal and optical measurement [97]. The MiniMed (Medtronic Diabetes, Northridge, California) Continuous Glucose Monitoring was an invasive device using a subcutaneous needle-type amperometric enzyme electrode based on $GO_X$ [98,99], which was attached on the abdomen of patient with a fixed belt. It acquired signal every 10 seconds and saved an average of the signal in memory every 5 min. Potts et al. [100] combined reverse iontophoresis extracting glucose from intact skin with an in situ glucose sensor in a device called the GlucoWatch G2 Biographer (GW2B, Cygnus, Redwood City, California) just like an electronic watch that was convenient for children to wear. Clinical researches including controlled clinical environment and the home environment had suggested the accuracy was relatively constant but not reliably in large blood glucose variation such as hypoglycemia [101, 102]. More current representative commercial CGM devices are provided in Table 1, with noninvasive method including ultrasonic, optical, electromagnetic, thermal and enzymatic sensing. It is observed that these devices are all small size, light weight and easy to wear. Further applications of the CGM devices may become a critical component of the closed loop insulin delivery system and, as such, must be selective, rapid, predictable and acceptable for continuous patient use [97-99]. However, three main accuracy limitations of CGM technology must be improved, including the calibration in rapidly changing condition [103, 104], the time lag because of blood-to-interstitial glucose transport and the sensor processing time [105, 106], as well as the error from random noise confound CGM data [107-109].

Table 1 Main reported commercial CGM devices with substantiated claims

| Brand name | Symphony | HG1-c | GlucoTrack | FreeStyle Navigator |
|---|---|---|---|---|
| Method | Ultrasound-based transdermal permeation | Raman spectroscopy | Ultrasonic, thermal and electromagnetic sensing | Enzyme-tipped catheter |
| Country | USA | USA | Israel | USA |
| Company | Echo Therapeutics | C8 MediSensors | Integrity Applications | Abbott |
| Website | http://echotx.com/ | http://www.c8medisensors.com/ | https://www.integrity-apps.com/ | http://www.abbott.com/ |
| Display | 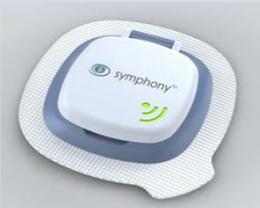 | 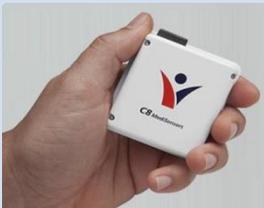 | 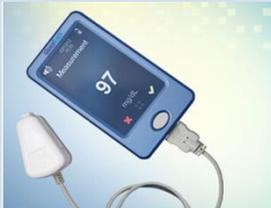 | 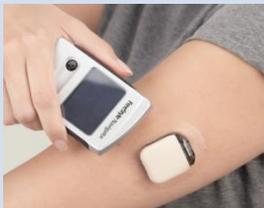 |
| Characteristic | Permeated skin measurement with some irritation [110,111], needing fast warm-up [112,113] | Scanning the skin with monochromatic source [114] and returning spectroscopy having high discrimination for glucose [115,116,188] | Attaching clip on the earlobe with abundant blood supply and not affected by activities [117,118] | Minimally invasive [119], needing warm-up about one hour and providing early-warning about high or low glucose [120] |

## 3.5 Pressure signal monitoring

Human body's pressure monitoring is an extensive concept, ranged from the obvious daily activity like the foot plantar pressure [121] to the subtle signal like the vibration of pulse [122].

In terms of daily activities monitoring, extensive research efforts have been made to access the accuracy of wearable sensors. Sazonov et al [123] developed an in-shoe pressure and acceleration sensor system for automatic recognition of postures and activities in patients with stroke, which could classify activities including sitting, standing and walking and detecting whether subjects were simultaneously performing arm reaching movements. Besides, combined wireless communication, the work undertaken by Bamberg et al. [124] had received much attention. Their device called GaitShoe developed complete wireless in-shoe system for gait analysis, which was capable of detecting heel-strike, estimating foot orientation and position.

On the other hand, with the development of flexible electronics, pressure sensors have emerged as a highly active field due to their unique advantages of flexibility, low-cost and high sensitivity. The mimicry of human skin to sense pressure is a topic of innovative research. Many physiological monitoring studies in this field have been made as shown in Fig. 3. Bao's team [125] developed a flexible pressure sensor based on organic thin film transistors and used it to monitor continuous radial artery pulse wave with no invasion and high fidelity. Pang et al. [126] also showed a flexible sensor that was based on two interlocked arrays of high-aspect-ratio Pt-coated polymeric nanofibres. This device has been used to measure the physical force of heart beat by attaching it directly above the artery of the wrist. Gong et al. [127] used ultrathin gold nanowires to develop wearable sensor, whose superior sensing properties in conjunction with mechanical flexibility and robustness enabled real-time monitoring of blood pulses. Other flexible systems were concentrated on human motion monitoring, such as the silk-molded pressure sensor for speaking recognition [128], the carbon nanotube strain sensor for movement and breathing [129], as well as the system including sensor, memory and actuator for diagnosis of movement disorders [130].

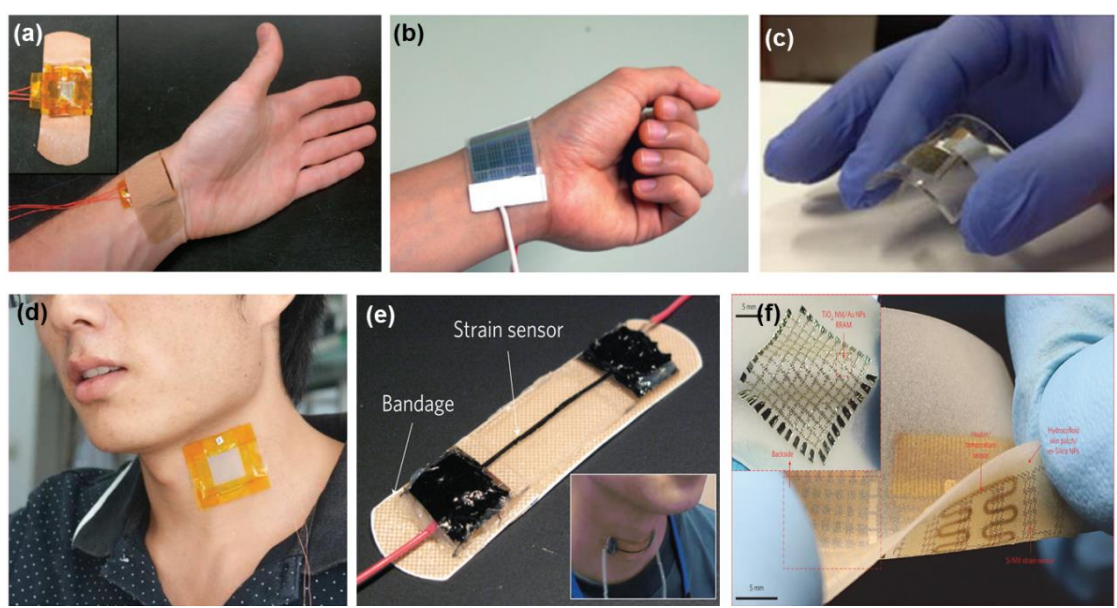

Fig. 3 Flexible electronics for various physiological measurement developed by different groups: (a) a flexible pressure sensor for continuous radial artery pulse wave [125]. (b) a flexible sensor for physical force of heart beat measurement [126]. (c) a flexible sensor for real-time monitoring of blood pulses [127]. (d) a silk-molded pressure sensor for speaking recognition [128]. (e) a strain sensor for movement and breathing monitoring [129]. (f) a system for movement disorder diagnosis [130].

## 4. Typical applications for chronic disease therapy

Based on the rapid development in monitoring systems for various physiological parameters, an emerging research trend of wearable system is to put more importance on developing therapy technique. Applications about wearable therapy devices with long-term monitoring can be divided into two systems, namely remote medical treatment and direct remedial treatment, while physiological parameters extracted from sensors can provide indicators of health status and have tremendous diagnostic value as well as therapeutic evaluations. Fig. 4 demonstrates a whole framework of remote medical treatment based on the monitoring of wearable sensors. These wearable sensors deployed according to the clinical application of interest are connected to mobile devices to gather different physiological information regarding the status of the individual. These measurements are forwarded via a wireless sensor network to a central connection node such as a mobile phone, then to a medical center. The professional physicians in medical center will give intervention to patients' behaviors based on the transmitted data, which can also be shared with families to keep acquaintance and deal with emergency situations whenever necessary. Systems for direct remedial treatment also rely on the wearable monitoring, but they take therapeutic measures directly like drug injection, defibrillation, auditory signal stimulation and more rather than relay suggestion from physician and behavior feedbacks remotely. Here, aiming at the chronic disease therapy, we will make a discussion on some typical wearable applications by dividing them into different functional assistance including organ, limb as well as daily habit management, and then review their respective developments and existing problems.

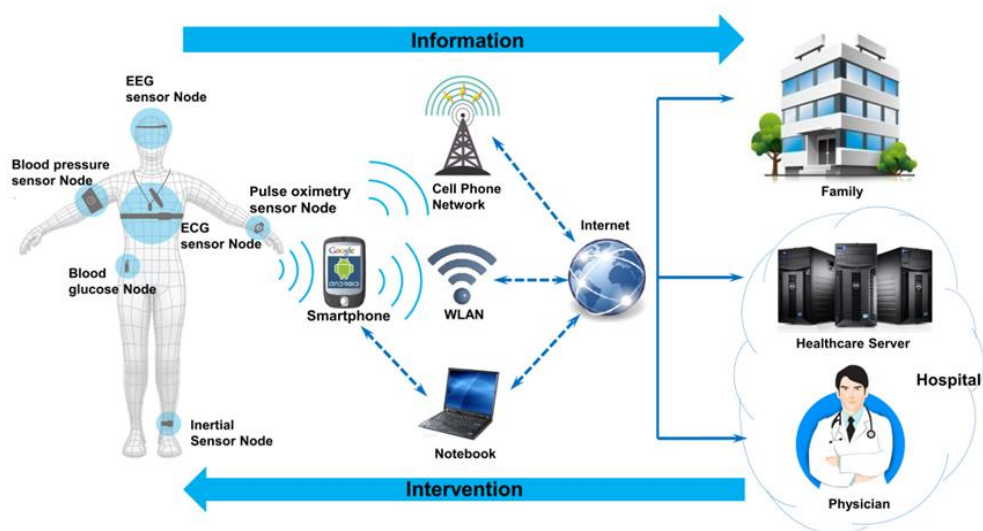

Fig. 4 Framework of telemedicine based on the monitoring of wearable sensors

## 4.1 Organ function assistance

The major hazard of chronic disease lies in the damage of vital organs like heart, pancreas and kidney. For patients with severe symptoms, the immediate and continuous function assistance of organ is very necessary when organs exhibit hormone imbalance or rhythm disorder. Wearable therapy devices for organ function assistance, which are also called as artificial organs, are aimed to timely response one or more abnormal conditions and maintain the homeostasis.

### 4.1.1 Wearable cardioverter-defibrillator for heart disease

Sudden cardiac death that mainly resulting from tachyarrhythmia is a major cause of mortality in industrialize populations, the rate of which in the United States, Netherlands, Ireland and China range from 50 to 100 cases per 100 000 in the general population annually [131]. After a few years of clinical investigation [132,133], the wearable cardioverter-defibrillator (WCD) that could provide an effective technology to improve survival rate of patients with a previous episode of sudden death, left ventricular dysfunction or ventricular tachyarrhythmias [134] was approved and introduced into practice until two years ago.

Currently, there are two WCD models available, the older LifeVest3100 and the new LifeVest 4000 [ZOLL Lifecor Corporation (ZOOL), Pittsburgh, PA, USA] [135]. Both of them used a chest garment to hold inplace two large posterior and one apical self-gelling defibrillationelectrodes, as well as four non-adhesive capacitivedry tantalum oxide electrodesmounted on an elastic belt for long-term ECG monitoring. The ECG-sensing electrodes provided two non-standardleads, front-back and left-right site bipolar ECG signals, for continuous electrocardiographic analysis. The three defibrillation electrodes were incorporated with a vibration plate to generate tactile notification as the first of a sequence of alarms once ventricular tachycardia or ventricular fibrillation was detected. However, the application of WCD useful for acute phase care is still limited according to the clinical practice [136] because it can only be a bridge to ICD (implantable cardioverter- defibrillator) or help avoid unnecessary ICD implantation.

### 4.1.2   Wearable closed-loop insulin pump for diabetes

Diabetes, with approximately 220 million diagnosed people in the worldwide [189], presents an indispensable problem to the health and social care systems. It is a chronic illness that requires continuing medical treatment and ongoing self-management to control the progression due to lack of permanent cure. In terms of continuing medical treatment for diabetes, insulin control of high blood glucose is the most effective means and the artificial pancreas that deliver extrinsic insulin automatically [137] has been proved a promising method.

The artificial pancreas (AP) is made up of CGM device and insulin pump while insulin delivery is modulated subcutaneously according to real-time interstitial glucose levels [138]. The major challenge in AP closed-loop system is refining the phase processes including overnight control, meals control and exercise control to avoid hypoglycemia. Now the simplest form of a closed-loop is to suspend insulin delivery at low glucose levels. Buckingham et al. [139] used five predictive algorithms to realize overnight control avoiding hypoglycemia in children and young adults with type 1 diabetes mellitus. In the clinical trial, insulin suspension was initiated at normoglycemia when the glucose trend was decreasing, and this measure prevented hypoglycemia on 75% of nights without a rebound to hyperglycemia. Del et al. [140] developed a new modular MPC algorithm focusing on meal control as the rapid rise of post-meal glucose is difficult to avert due to the inherent delays in subcutaneous insulin absorption and action, which showed satisfactory dinner control versus open-loop. This system running on a smart-phone in outpatient conditions reduced hypoglycemia but accompanied with marginal increase in average glycemia resulting from a possible overemphasis on hypoglycemia safety [141]. Co-administration of insulin and glucagon was investigated by El-Khatib et al. [142] to control the blood glucose at normal range. No hypoglycemia was observed during a 24h period of day-and-night closed-loop delivery in 11 adults once an appropriate pharmacokinetic model of insulin absorption was used. The MiniMed Paradigm® REAL-Time system was launched as early as 2006, which integrated real-time CGM with an insulin deliver device [143]. It was small in size so patients could wear it almost anywhere such as under the clothing in a leg pouch or on the belt like a mobile. Insulin delivery and CGM were made possible through tiny components patients wear for up to six days at a time [190]. Although the REAL-Time system did not automatically adjust insulin delivery based on sensor glucose data in its current embodiment, this integrated system formed the basic platform for further generations of products in which the sensor will modify insulin dosing in semi- and ultimately fully closed-loop modalities.

### 4.1.3 Wearable artificial kidney for chronic kidney disease

A wearable artificial kidney (WAK) is a long-held aim in the treatment of patients with chronic kidney disease, provided that it would combine continuous blood purification, preventing the fluctuations and complications in intermittent dialysis for three time a week normally [144,145]. Though the WAK is not commercialized owing to some technical problems, numerous efforts has been carried out for decades and the breakthroughs and the issues are summarized in Table 2. Treatment modalities of dialysis includes hemodialysis (HD), hemofiltration(HF) and peritoneal dialysis (PD), while all three need to minimize the huge dialysate supply system to realize wearability and portability [146]. Pioneers proposed the concept that continuous regeneration and reuse of a small amount of dialysate in a closed-loop system instead of using large volumes of dialysate in a single pass configuration as done in conventional dialysis. Kolff et al. [147,148] provided a portable WAK system consisted of pumps, batteries, dialyzers and a charcoal regeneration cartridge, where 20 l of dialysate tank was attached for 1.5-2h for urea and potassium removal. Due to large dialysate volume for electrolyte balance, it was portable rather than wearable. Murisasco et al. [149,150] developed a more advanced regeneration module based on the combination of sorbents and enzyme which was called recirculating dialysis (REDY) system. This cartridge used for purifying the dialysate included urease for removal of urea, charcoal for absorption of non-urea organic toxins, zirconium phosphate for removal of potassium and ammonium, and zirconium oxide for removal of phosohate. However, a large amount of zirconium phosphate required to remove the generated ammonium limits further miniaturization. After that, more different approaches were tried. Gura et al. [151,152] developed a belt-type wearable HD machine that can accomplish dialysate regeneration with 375 ml of dialysate volume and got acceptable clinical results. However, this device had some limitations such as inaccurate electrolytes, acid–base control and ammonium accumulation in the sorbent cartridge. It is to be observed that the PD as a blood-free dialysis technique can be an alternative to HD and HF approaches. Roberts et al. [153] proposed a concept of tidal PD using a hemofilter and REDY sorbent cartridge for WAK. But at that time, an appropriate double-lumen PD catheter was not available and the safety of sorbent cartridge was not properly verified in clinical. Ronco et al. [154] showed the concept of a wearable device called ViWAK based on sorbent cartridge technology and double-lumen PD catheter. Although this device was not applied in clinical trials, the main innovation was the remote control with a handheld computer. The common disadvantage of PD was an addition of an injection system for glucose and bicarbonate to maintain acid-base balance and osmotic force.

Table 2 Historic view on wearable artificial kidney

| Name | TM | Technology breakthrough | Issues | Clinical Experiment | Years | Ref. |
|------|-----|------------------------|--------|---------------------|-------|------|
| Portable WAK | HD | A charcoal regeneration cartridge | Large dialysate volume | 5 patients, 3h , 6 times per week | 1976 | [147,148] |
| Continuous Hemofiltratiom | HF | Sorbent and enzyme technology (REDY) | Large amount of zirconium phosphate | 2 patients, one for 1 month and another for 1 month | 1986 | [149,150] |
| Flow-thru PD | PD | Concept of tidal PD, REDY sorbent cartridge | Additional injection system for glucose and bicarbonate | Before clinical | 1999 | [151] |
| The WAK | HD | Pulsatile pump, REDY sorbent cartridge | Variation of composition and PH in dialysate | 8 patients for 4-8 h | 2007 | [152] |
| ViWAK | PD | Remote control, double lumen PD catheter | Variation of composition and PH in dialysate, glucose injection | Before clinical | 2007 | [153] |
| The WHF | HF | Pulsatile pump, improved middle molecule clearance | Low small solute clearance | 6 patients for 6h | 2008 | [155] |

## 4.2 Limb function assistance

The aim of wearable therapy devices applied on limb function assistance is to facilitate the patient's task performance after a neurological injury. Without these wearable devices, training based on manual assistance is exhausting for the staff, and the duration of training is thus limited by the fatigue of the therapist. Besides, the manual assistance is lack of the repeatability of training and quantitative evaluation of curative effect. These disadvantages can be relieved or even overcome with the use of wearable therapy devices.

### 4.2.1 Therapy of Parkinson's disease based on symptoms monitoring

Parkinson's disease is a common neurodegenerative disorder caused by a deficiency of dopamine due to the severe neuronal dysfunction in the substantia nigra pars compacta [156], which affects about 3% of the population over the age of 65 years and more than 500,000 U.S. residents [157]. Current therapy is mainly based on pharmacotherapy using the biosynthetic precursor levodopa or drugs to activate dopamine receptors, which just maintains effective for some time but eventually develop complications like motor fluctuations and psychiatric disturbances.

Another key factor reducing the patients' quality of life is motion problems such as tremor, freezing of gait, postural instability. It is well recognized that auditory, visual or tactile stimuli can improve gait in patients with Parkinson's disease. As a result, devices that exploit the sensory cueing-related modification of gait in patients have been proposed. Bachlin et al. [158] developed a wearable system with acceleration sensors to detect freezing of gait (FOG) and provided a rhythmic auditory signal to stimulate the patients to resume walking. The signal that 1Hz ticking sound starting whenever a FOG episode is identified and ending when the patient resumes walking, verified that discontinuous and selective external cueing can keep the effect to improve the gait stability. Ferrarin et al. [159] demonstrated a pair of optical stimulating glasses that provided different types of continuous optic flow and intermittent stimuli synchronous with external events. After prolonged testing, the device showed a remarkable increase in stride length for all subjects. Espay et al. [160] examined the efficacy of a closed-loop wearable visual-auditory cueing device in 13 patients with Parkinson's disease with off-state gait impairment. After 2 weeks of twice daily use at home, an overall improvement in gait was measured.

### 4.2.2  Wearable rehabilitation robots in neurologic injuries

Wearable robots known as robotic exoskeleton and powered orthoses are being developed for movement disabilities caused by neurologic injuries like stroke, traumatic brain and spinal cord injuries [161,162]. Positive outcome of physical rehabilitation depends heavily on functional task-oriented training that can be realized by mechanical and electrical devices with high-intensity and repetitive training.

Traditionally, actuators of the devices are provided in three forms: electric current, hydraulic fluid or pneumatic pressure. The electrical actuators are most common because of their ease to provide and store electrical energy as well as their relatively higher power. Hasegawa et al. [163] introduced an exoskeleton assistive hand that supports human hand and wrist activities by using user's bioelectric potential to control movement. Rocon et al. [164] proposed an exoskeleton for tremor assessment and suppression with DC motors actuator as shown in Fig. 5(a). Powered orthosiswith electric motors controlled by joint angle were also developed to assist elbow rehabilitation or foot drop gait [165,166]. Pneumatic artificial muscle was used in upper extremity therapy robot [167] and soft robotic device for ankle foot rehabilitation [168], while its weight was generally light compared to other actuators but also have slow and non-linear dynamic response. Fig. 5(b) demonstrated the prototype of soft robotic device for ankle foot rehabilitation.

Miniaturized and flexible fluidic actuators were applied in the elbow orthosis proposed by Pylatiuk et al. [169]. Moreover, another specific electrical stimulation known as functional electrical stimulation (FES) is the natural actuator of body muscles instead of traditional external actuators. From a therapeutic point of view, FES allows patients to exercise muscle and preventing muscular atrophy [170]. However, disadvantages are also related to the appearance of muscle fatigue and the control of joint trajectories [171]. Two systems combining FES with assistive force were proposed by Freeman et al. [172] and Li et al. [173]. There were also some commercial systems using FES for limb rehabilitation: H200 (Bioness, US) for hand paralysis, L300 (Bioness, US) for foot drop and NeuroMove (Zynes Medical, US) for stroke. Fig. 5(c) demonstrated the Bioness H200 system [191] including the stimulator and the control unit and the procedure that using this device to realize grasp. Patients could control stimulation to their hand by using the unaffected hand to press buttons on the separate control unit. The press of a button activated hand extensors and another button press turned off extensors and activated flexors.

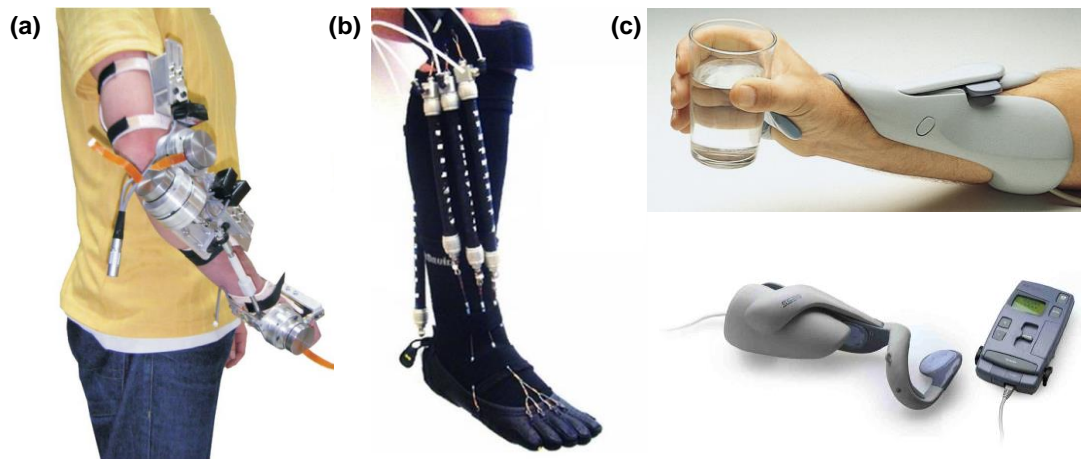

Fig. 5 Wearable robotic exoskeleton and powered orthoses for limb rehabilitation: (a) wearable orthosis with electrical actuator [164]. (b) wearable robotic device with pneumatic actuator [168]. (c) The Bioness H200 based on FES as natural actuator of body muscles [191].

## 4.3 Daily habit management assistance

Patients with chronic health conditions struggle with multifaceted needs that change during the course of their disease journeys. One path for the development of wearable devices may lie in patient-driven health care, which encourages greater patient-physician collaboration, expands patient social networks and increases patient use of personal data for tracking their health status. Through establishing the validity of data gathering, wearable monitoring devices can make a direct and real-time impact on daily habit management assistance. Here, taking diabetes as an example, we will discuss the specific management details and the latest advances.

The diabetes' self-management tasks such as blood glucose monitoring, physical exercise and diet are inherently data intensive, requiring acquisition, storage and analysis of large amounts of data on a regular basis [174]. One way that has been proposed to help individuals better manage their diabetes is to make full use of smart-phone technology, which is coupled with the proliferation of data connectivity and relatively inexpensive cost. El-Gayar et al. [175] reviewed 16 articles and 71 applications and divided the features of app functions into two classes. The primary features included blood glucose monitoring, medication, diet management and physical exercise. To form a long-term record and get professional analysis from health care provider, some apps needed patients enter manually into the mobile phone and others realized data entry automation just for blood glucose through blue-tooth [176]. The second features included weight, blood pressure, education, alert, decision support and social networking. It was important to note that decision support which mainly referred to automated analysis and ruled-based interpretation of data on an individual basis was a comprehensive but crucial part of diabetes management. However, only one of the 71 applications that provided decision support had Food and Drug Administration (FDA) clearance in order to provide safety and assurance. DiabeteManager coming from WellDoc™ initially, which was the only one example with FDA approval, had both clinical and marketable validation. WellDoc published some clinical results that documented the positive impact of their multipronged approach to patient and physician satisfaction [177,178].

Last year, the company launched new paid application called BlueStar which received approval for reimbursement as a pharmacy benefit for employees of several large companies including Ford Motor Company, RiteAid and Dexcom [179]. The BlueStar platform still required a physician prescription which could reduce the cost of physician-patient communication and partly account for the trust of patients on the support decision. In general, management system based on mobile phone is designed to make patients more acquainted with their own physical conditions and more responsible for their own day-to-day care.

## 5. Conclusions

The aim of this article is to provide an overview on current status and future prospects about the research and development of wearable devices related to chronic diseases therapy. For this goal, it is necessary to define the field of wearable systems. These systems can be divided into two categories, namely long-term monitoring of physiological signals, as well as continuous therapy including organs, limbs and daily habits assistance.

Long-term or continuous wearable monitoring systems with unobtrusive sensors took in many advanced technologies like the electronic circuit, MEMS and nanotechnology, as well as novel approaches like noncontact ECG and glucose monitoring through blood substitutes. On this basis, information technology expedited the treatment with pharmaceutical products becoming more accurate with continuous automatic processes to dispense an effective dose of drug to treat disease, whose ultimate target is to replace defective organs. In the meantime, we have also witnessed a growing interest for the emerging need for establishing a telemedicine in the home setting to realize self-management and implement clinical interventions. Besides, the scope of chronic disease treatment through wearable devices is broadened as the miniaturization of therapy equipment and clinical rehabilitation evidences with attempting various remedy methods.

However, most systems are still in their prototype stages and there remain many problems to be solved. Firstly, battery technologies appears to be perhaps the biggest technical issue and performance bottleneck in current wearable devices that should maintain operating for long periods and avoid replacing and charging frequently. Low-power transceivers and power scavenging are supposed to be developed as quickly to reduce energy consumption. Secondly, security of private information about the patient's health status must be taken into account. Proper encryption and identity verification are needed to ensure the privacy of all communicated data in telemedicine. Lastly, clinical experiments to prevent infection and inflammation for some invasive or minimal invasive wearable systems should be done. Developed systems must be exhaustively tested and validated by professional physicians.

There is plenty of room for further technological developments in wearable systems. Here, we envision the following emerging directions for wearable monitoring and therapy devices worthy of being emphasized in the near future.

First of all, flexible electronics, where electronic circuits are manufactured on flexible substrates such as paper, cloth fabrics and directly on human body to provide sensing, powering and interconnecting functions, have a broad range application like smart textile and transcutaneous film as new wearable monitoring systems. It is worth mentioning that, rapid advances in this field promise to allow one to print a full circuit on fabric or skin and new materials like liquid metal [180] have the potential to directly print on skin to take effect. In addition, tissue engineer technology could contribute to mimicking physiological function of body organs. Researchers have suggested the concept of a renal assist device with cartridges containing human tubular cells [181,182], more studies should be done in vitro or vivo to enlarge the function of wearable even implantable system. In the end, business models of wearable monitoring and therapy devices need to be established. Only the wearable devices are used and accepted by patients as a reliable, multifunctional and easy-to-use technology, can they relieve the severe burdens of medical system and go farther in the avenue of improving quality of life.

## Acknowledgements

This work is partially supported by The Joint Research Fund from Ministry of Higher Education, NSFC Key Project under Grant No. 91748206, as well as Dean's Research Fund and the Frontier Project of the Chinese Academy of Sciences.